\RequirePackage{lineno}
\documentclass[twocolumn,letterpaper,aps,prc,longbibliography,superscriptaddress,nofootinbib,floatfix]{revtex4-1}

\usepackage{graphicx}	
\usepackage{appendix}
\usepackage{float}
\usepackage{bbm}
\usepackage{hyperref}

\hypersetup{
    colorlinks=true,
    linkcolor=blue,
    filecolor=blue,      
    urlcolor=blue,
    citecolor=blue,
}

\usepackage{xspace}	

\newcommand{\ecc}{\mbox{$\varepsilon$}\xspace}
\newcommand{\pt}{\mbox{$p_T$}\xspace}
\newcommand{\npart}{\mbox{$ N_{\rm part} $}\xspace}
\newcommand{\ncoll}{\mbox{$ N_{\rm coll} $}\xspace}

\newcommand{\dnchdeta}{\mbox{$dN_{\rm ch}/d\eta$}\xspace}
\newcommand{\mdnchdeta}{\mbox{$\langle dN_{\rm ch}/d\eta\rangle$}\xspace}

\newcommand{\sqsn}{\mbox{$\sqrt{s_{_{NN}}}$}\xspace}

\newcommand{\pp}{\mbox{$p$$+$$p$}\xspace}
\newcommand{\dau}{\mbox{$d$$+$Au}\xspace}

\newcommand{\pau}{\mbox{$p$$+$Au}\xspace}
\newcommand{\pa}{\mbox{$p$$+$$A$}\xspace}

\newcommand{\pO}{\mbox{$p$$+$O}\xspace}
\newcommand{\pPb}{\mbox{$p$$+$Pb}\xspace}
\newcommand{\ppb}{\mbox{$p$$+$Pb}\xspace}
\newcommand{\OO}{\mbox{O$+$O}\xspace}
\newcommand{\heauthree}{\mbox{$^{3}$He$+$Au}\xspace}
\newcommand{\heaufour}{\mbox{$^{4}$He$+$Au}\xspace}

\newcommand{\beau}{\mbox{$^{7,9}$Be$+$Au}\xspace}
\newcommand{\oau}{\mbox{O$+$Au}\xspace}
\newcommand{\cau}{\mbox{C$+$Au}\xspace}

\newcommand{\xexe}{\mbox{Xe$+$Xe}\xspace}
\newcommand{\pbpb}{\mbox{Pb$+$Pb}\xspace}

\newcommand{\sonic}{\mbox{{\sc sonic}}\xspace}

\newcommand{\ampt}{\mbox{{\sc ampt}}\xspace}

\begin{document}



\title{Exploring New Small System Geometries in Heavy Ion Collisions}

%

\newcommand{\colorado}{University of Colorado, Boulder, Colorado 80309, USA}
\newcommand{\saclay}{CEA/IPhT/Saclay, Orsay, 91190, France}
\newcommand{\ornl}{Oak Ridge National Laboratory, Oak Ridge, Tennessee 37831, USA}
\newcommand{\frib}{Facility for Rare Isotope Beams, Michigan State University, East Lansing, Michigan 48824, USA}
\newcommand{\lanl}{Los Alamos National Laboratory, Los Alamos, New Mexico 87545, USA}
\newcommand{\tud}{Institut f$\ddot{u}$r Kernphysik, Technische Universit$\ddot{a}$t Darmstadt, 64289 Darmstadt, Germany}
\newcommand{\gsi}{ExtreMe Matter Institute EMMI, GSI Helmholtzzentrum f$\ddot{u}$r Schwerionenforschung GmbH, 64291 Darmstadt, Germany}

\author{S.H.~Lim} \affiliation{\colorado}
\author{J.~Carlson} \affiliation{\lanl}
\author{C.~Loizides} \affiliation{\ornl}
\author{D.~Lonardoni} \affiliation{\frib} \affiliation{\lanl}
\author{J.E.~Lynn} \affiliation{\tud} \affiliation{\gsi}
\author{J.L.~Nagle} \affiliation{\colorado}\affiliation{\saclay}
\author{J.D.~Orjuela Koop} \affiliation{\colorado}
\author{J.~Ouellette} \affiliation{\colorado}














\date{\today}

\begin{abstract}

Relativistic heavy ion collisions produce nuclei-sized droplets of quark-gluon plasma whose expansion is well described by viscous hydrodynamic calculations. Over the past half decade, this formalism was also found to apply to smaller droplets closer to the size of individual nucleons, as produced in $p$$+$$p$ and $p$$+$$A$ collisions.
The hydrodynamic paradigm was further tested with a variety of collision species, including $p$$+$Au, $d$$+$Au, and $^{3}$He$+$Au producing droplets with different geometries.    Nevertheless, questions remain regarding the importance of pre-hydrodynamic evolution and the exact medium properties during the hydrodynamic evolution phase, as well as the applicability of alternative theories that argue the agreement with hydrodynamics is accidental.   In this work we explore options for
new collision geometries including $p$$+$O and O$+$O proposed for running at the Large Hadron Collider, as well
as, $^{4}$He$+$Au, C$+$Au, O$+$Au, and $^{7,9}$Be$+$Au at the Relativistic Heavy Ion Collider.
\end{abstract}

	



\maketitle



\section{Introduction}
The standard evolution picture for relativistic heavy ion collisions includes the formation of a quark-gluon plasma 
that flows hydrodynamically as a nearly perfect fluid~\cite{Romatschke:2017ejr,Heinz:2013th}.   Smaller collision
systems, including \pp and \pa at the Relativistic Heavy Ion Collider (RHIC) and the Large Hadron Collider (LHC), reveal
particle emission patterns consistent with this standard evolution picture---for recent reviews see Refs.~\cite{Nagle:2018nvi,Loizides:2016tew}.   Specific experimental tests with incoming nuclei of different intrinsic geometries have been
proposed to further test this picture and also provide additional constraints on pre-hydrodynamic initial conditions and evolution, as well as on the medium properties during hydrodynamic evolution. A specific hypothesis was put forward with
predictions for elliptic and triangular flow patterns in \pau, \dau, \heauthree collisions at 
RHIC~\cite{Nagle:2013lja}.  Experimental data from the PHENIX experiment in all three systems and for both flow patterns are quite well described within this framework~\cite{Aidala:2018mcw}.

RHIC has the unique ability to collide, essentially, any pair of ions, even of different nuclear species.   Specific proposals for colliding
deformed uranium nuclei~\cite{Voloshin:2010ut} and comparing the collisions of different isobars~\cite{Skokov:2016yrj} (Zr/Ru) have been implemented at RHIC to test predictions for the chiral magnetic effect and other magnetic field-correlated effects. There have even been proposals that posit that the hydrodynamic evolution
is so well understood that one can test hypotheses about possible alpha clustering in small nuclei such as beryllium,
carbon, and oxygen via their imprint in the initial geometry~\cite{Rybczynski:2017nrx,Zhang:2017xda,Bozek:2014cva}.

Unlike RHIC, the LHC has only one ion source and is thus limited to symmetric collisions of nuclei, for example, \pbpb, and asymmetric systems such as \pa but where one of the beams is necessarily a proton beam.   
Opportunistically, for accelerator-related studies, the LHC recently had a run of \xexe lasting only a few hours.    Despite the collision system not being specifically motivated for testing a particular hypothesis, the 
run provides additional data---see, for example, Ref.~\cite{Acharya:2018ihu}, that overall confirms the standard hydrodynamic evolution picture~\cite{Giacalone:2017dud} with a couple of
interesting puzzles.  In a similar way, there is now an exciting possibility 
for colliding proton on oxygen (\pO) and oxygen on oxygen (\OO) in upcoming LHC running~\cite{Citron:2018lsq}.

Despite the wealth of data in small systems being described quantitatively by the standard hydrodynamic
picture---see, for example, Ref.~\cite{Weller:2017tsr},---there are completely orthogonal calculations that produce
correlations purely from the initial scattering via color fields (see Ref.~\cite{Dusling:2015gta} for a useful review).
Recently, one variant of such color field calculations was able to qualitatively describe the ordering of elliptic flow in \pau, \dau, \heauthree~\cite{Mace:2018vwq}, though the reasoning for the ordering is not fully understood~\cite{Nagle:2018ybc}.   A critical path to follow to resolve these questions is to have all models
provide documented, publicly available code and then to see whether further discrimination and improved understanding
can be gleaned from existing experimental data.   At the same time, it is worth asking whether additional data in
new collision systems can also be elucidating.    

One such proposal for new data is to collide polarized deuterons on large nuclei~\cite{Bozek:2018xzy} and thus to have
information on the orientation of the deuteron constituent proton and neutron.    This clever idea would
specifically test that the correlations are established with respect to the initial geometry of the collision, which would
presumably not be true in the context of color field descriptions.    
At the top RHIC energy, accelerating polarized deuterons in the full ring is a problem because the Siberian snake  magnets~\cite{Roser:2002yz} are
ineffective due to the small anomalous magnetic moment of the projectile\footnote{Private communication from Wolfram Fischer}.
Potential collisions in the fixed-target mode of the Pb beam with a polarized deuteron target is another option that could
be explored with LHCb.    Another proposal is to run
\heaufour collisions at RHIC, where the more compact projectile might have measurable effects relative to
existing data with \heauthree\footnote{Communicated to one of us by Berndt M\"{u}ller}.

In this study, we explore various new collision geometries in the context of the publicly available 
hydrodynamic model \sonic~\cite{Habich:2014jna,Romatschke:2015gxa}.    
We have used a shear viscosity to entropy density ratio of $\eta/s = 0.08$, and bulk viscosity ratio of $\zeta/s=0$ in our \sonic calculations.
We have incorporated full $N$-nucleon
configurations for light nuclei including $^{4}$He, carbon, and oxygen and then used a Monte Carlo
Glauber calculation for the initial conditions.  

\section{Modeling Light Nuclei Configurations}

The modeling of the initial collision between nuclei is carried out within the context of
the Monte Carlo Glauber framework~\cite{Miller:2007ri}.  In such calculations, the nucleons in each nucleus
are generally randomly distributed according to a Woods-Saxon distribution such that---for a given impact parameter
between the two nuclei---one can determine whether individual nucleon-nucleon pairs from the projectile and target
experience elastic or inelastic collisions.    For previous studies, full three-nucleon coordinate spatial configurations were
calculated for $^3$He~\cite{Nagle:2013lja} and made publicly available as part of the PHOBOS Monte Carlo Glauber package~\cite{Loizides:2014vua}.   Calculations in a similar formalism are now included for four-nucleon
configurations of $^4$He.   
These configurations are generated from a Green’s Function Monte Carlo simulation of $^4$He using the Argonne V18 two-nucleon~\cite{Wiringa:1994wb} and Urbana IX~\cite{Pudliner:1997ck} three-nucleon potentials. The resulting binding energy and charge radius are in very good agreement with experiment (better than 1\% agreement)---see Ref.~\cite{Carlson:2014vla} for more details.

Full 12-nucleon configurations are included for carbon---see Ref.~\cite{Carlson:2014vla} for details.   Full 16-nucleon configurations are calculated for oxygen.
The $^{16}$O configurations are generated from quantum Monte Carlo calculations using chiral effective field theory Hamiltonians. 
In more detail, the employed interaction consists of two- and three-body local chiral potentials at next-to-next-to-leading order (N$^2$LO) with 
the $E\mathbbmss{1}$ parametrization of the three-body contact term and coordinate-space cutoff $R_{0}=1.0$~fm, see Refs.~\cite{Gezerlis:2013ipa,Gezerlis:2014,Tews:2015ufa,Lynn:2016,Lynn:2017,Lonardoni:2018prl,Lonardoni:2018prc}.
The initial trial wave function is optimized to minimize the variational expectation value of the chiral Hamiltonian~\cite{Lonardoni:2018prl,Lonardoni:2018prc}. Such an optimized variational wave function is then used to extract the nucleus configurations in coordinate space. 

\section{Initial Conditions}

With the nucleon configurations for the relevant nuclei in hand, we calculate the initial energy density 
deposited in the collision of two nuclei in the two-dimensional transverse plane.    There are various
methodologies for such calculations, the simplest of which is to include a two-dimensional Gaussian
distribution of energy for each participating nucleon (i.e.,\ any nucleon with at least one inelastic collision).
As the simplest case, we employ a Gaussian of width $\sigma = 0.4$~fm and then have a single scale factor on the overall
energy density such that, when run through the full hydrodynamic evolution and hadronic cascade of the \sonic model, 
one matches the measured (or predicted) charged particle multiplicity (\dnchdeta).    

For small system collisions at \sqsn = 200 GeV at RHIC, a nucleon-nucleon inelastic cross section of 42~mb is employed, and the scale factor is determined so as to match the measured \dnchdeta at midrapidity in \heauthree collisions
measured by the PHENIX experiment~\cite{Adare:2018toe}.   The identical parameter set, including this scale factor, is
utilized for all the RHIC collision system predictions.  For the \pO and \OO calculations at \sqsn=~7 TeV at the LHC, a nucleon-nucleon inelastic cross section of 71~mb is employed, and the scale factor is determined by 
matching the measured \dnchdeta at midrapidity in \ppb collisions at \sqsn=~5.02 TeV and 8.16 TeV~\cite{Acharya:2018egz,Sirunyan:2017vpr,ALICE:2012xs}. A single scale factor with different inelastic cross sections, namely 68~mb for \sqsn=~5.02 TeV and 72~mb for \sqsn=~8.16 TeV, can reproduce \dnchdeta at both energies. Of course, in these systems for which no measurements exist, an important check before comparing flow predictions will be to compare particle multiplicity distributions with the
measurements after data taking.

As an alternative calculation for the energy density, we utilize the publicly available IP-Jazma framework~\cite{Nagle:2018ybc}.   The calculation follows the observation that eccentricities from
a full calculation in the saturation dense-dense limit, as encoded in IP-Glasma~\cite{Schenke:2012wb}, 
are fully reproduced by modeling Gaussian distributions for each participating nucleon 
(motivated by the IP-Sat model~\cite{Kowalski:2003hm}) and then simply multiplying the
target and projectile densities in lattice sites in the transverse plane~\cite{Romatschke:2017ejr}.   
This modeling should provide a closer match than the TRENTO model with parameter $p=0$, often used to
approximately match IP-Glasma conditions~\cite{Moreland:2014oya}.    We note that neither model has the single lattice site spikes characteristic of IP-Glasma calculations, which are essentially artifacts regardless.
We have run the IP-Jazma code in
the dense-dense mode, without $Q_{s}^{2}$ fluctuations, and with IP-Sat Gaussian width 0.32~fm.
These IP-Jazma initial conditions should lead to more compact hot spots in \OO collisions and thus to somewhat steeper pressure gradients.   Of course, these
can be reduced by including an additional pre-hydrodynamization stage---often modeled via free streaming.
For the central \pO case, the IP-Jazma initial conditions would lead to a somewhat circular energy density distribution, just as earlier
implementations of IP-Glasma, and including an \textit{ad hoc} three-quark substructure would mitigate that effect~\cite{Mantysaari:2017cni}.    For now, we leave such modeling of nucleon sub-structure to future calculations.

\section{Predictions:   Oxygen Systems \pO and \OO}

\begin{figure*}[htb]
\includegraphics[width=1.00\linewidth]{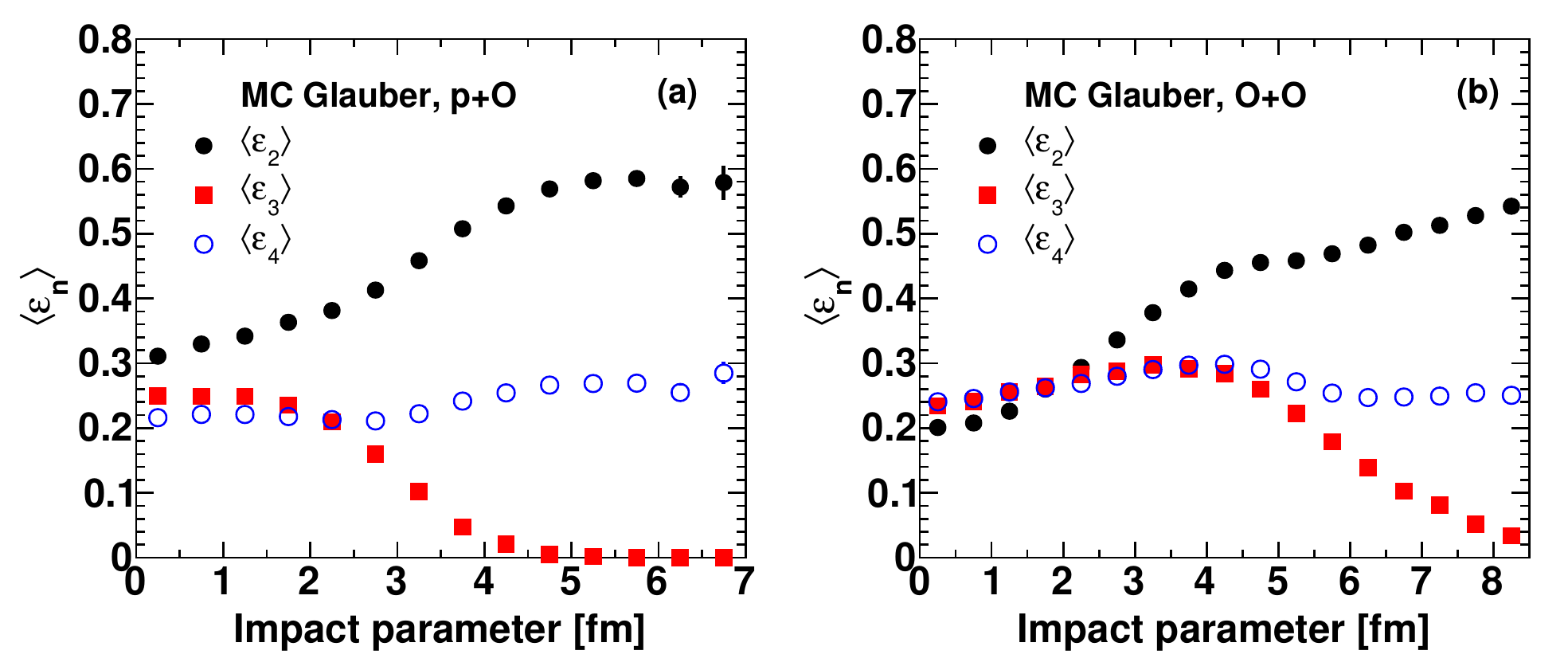}
\caption{\label{fig:ecc_pO_OO}
Mean eccentricity ($\langle\ecc_{2}\rangle,~\langle\ecc_{3}\rangle,~\langle\ecc_{4}\rangle$) as a function of impact parameter in \pO and \OO collisions at $\sqsn=7~{\rm TeV}$.}
\end{figure*}

\begin{figure*}[htb]
\includegraphics[width=1.00\linewidth]{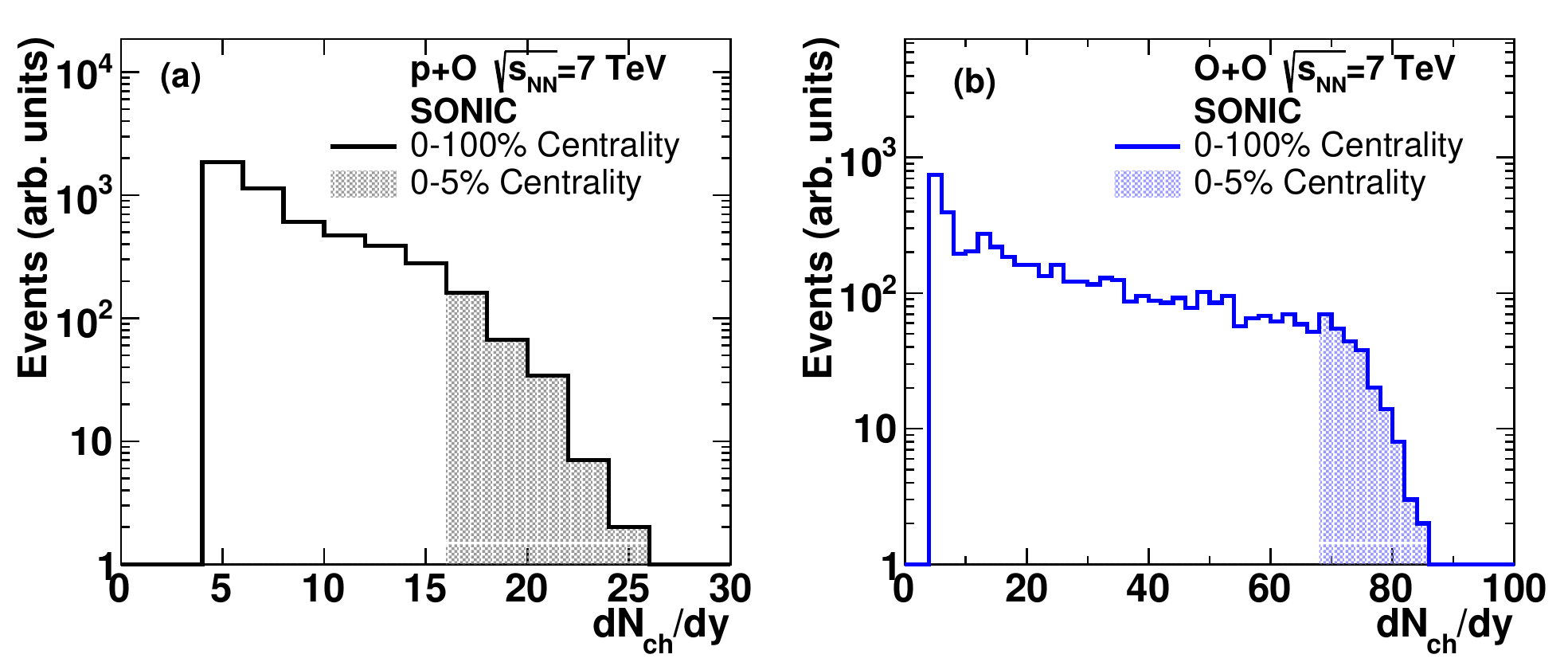}
\caption{\label{fig:mult_sonic}
\dnchdeta at midrapidity from \sonic in unbiased (0\%--100\%) \pO and \OO collisions at $\sqsn=7~{\rm TeV}$, and the shaded region is corresponding to 5\% events of highest \dnchdeta.}
\end{figure*}

\begin{figure*}[htb]
\includegraphics[width=1.00\linewidth]{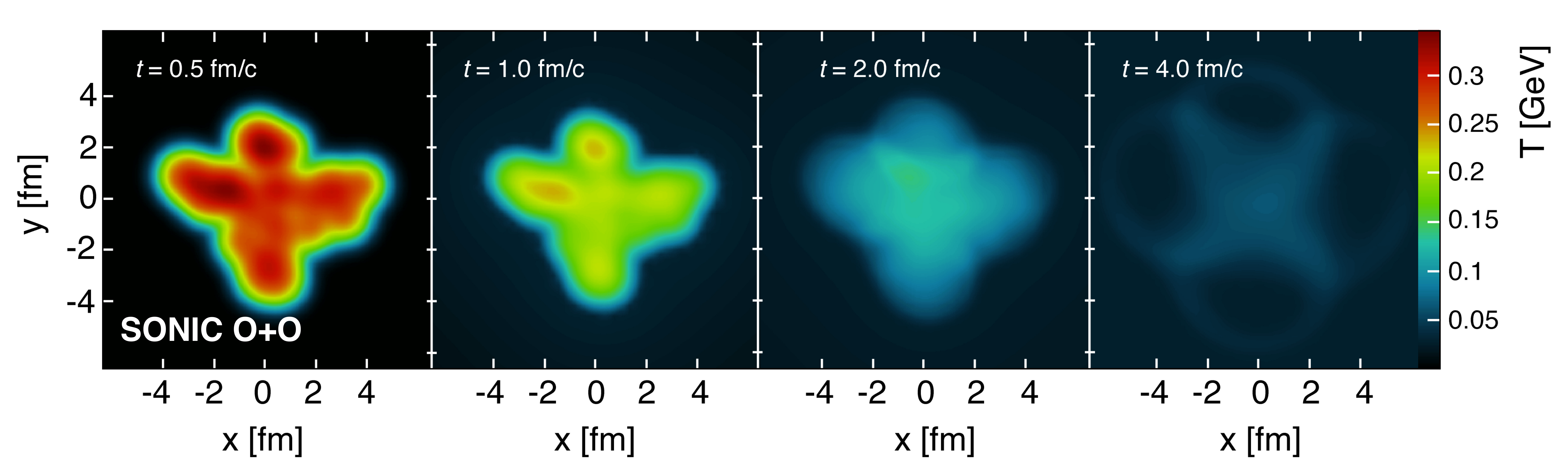}
\caption{\label{fig:event_OO}
An example of time evolution of a \OO event from \sonic; the color scale indicates the local temperature.}
\end{figure*}

\begin{figure*}[htb]
\includegraphics[width=1.00\linewidth]{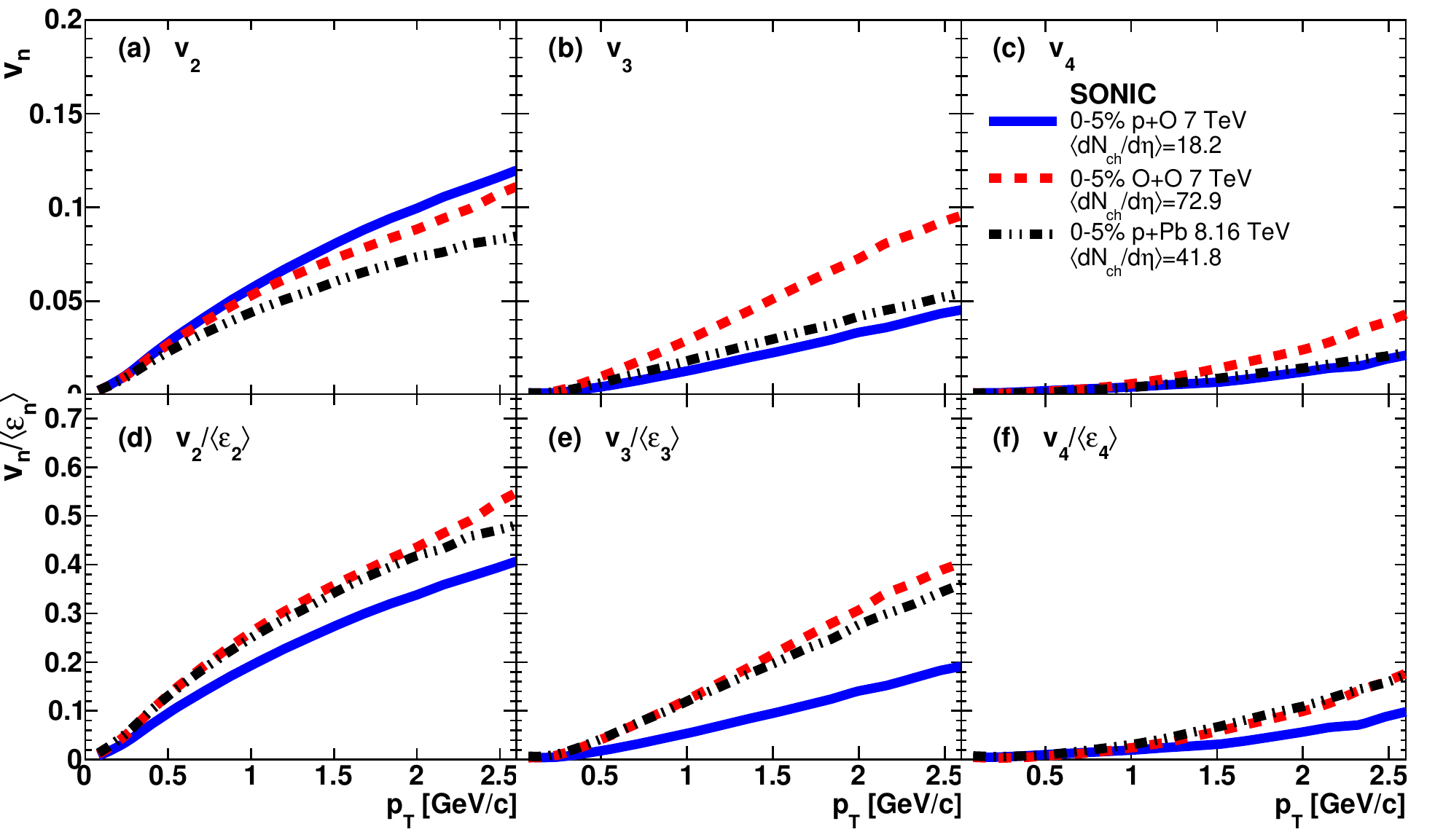}
\caption{\label{fig:vn_pt_pO_OO_pPb_0005}
Comparison of $v_n$ and $v_{n}/\langle\ecc_{n}\rangle$ as a function of \pt in 0\%--5\% of three collision systems, \pO and \OO collisions at $\sqsn=7~{\rm TeV}$ and \pPb collisions at $\sqsn=8.16~{\rm TeV}$.}
\end{figure*}

\begin{figure*}[htb]
\includegraphics[width=1.00\linewidth]{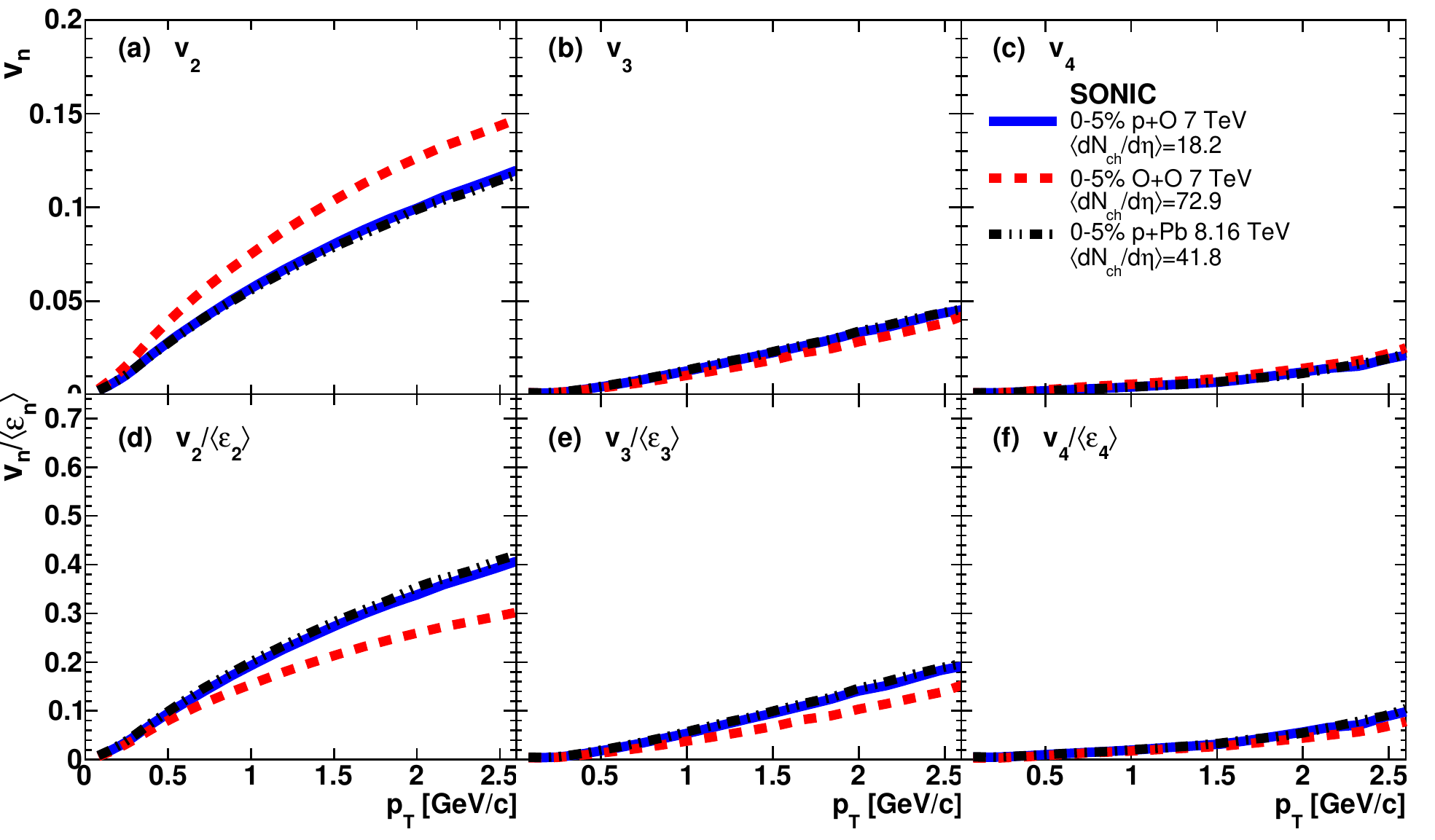}
\caption{\label{fig:vn_pt_pO_OO_pPb_match}
Comparison of $v_n$ and $v_{n}/\langle\ecc_{n}\rangle$ as a function of \pt in a 5\% centrality range of collisions showing a similar \mdnchdeta.}
\end{figure*}

\begin{figure*}[htb]
\includegraphics[width=1.00\linewidth]{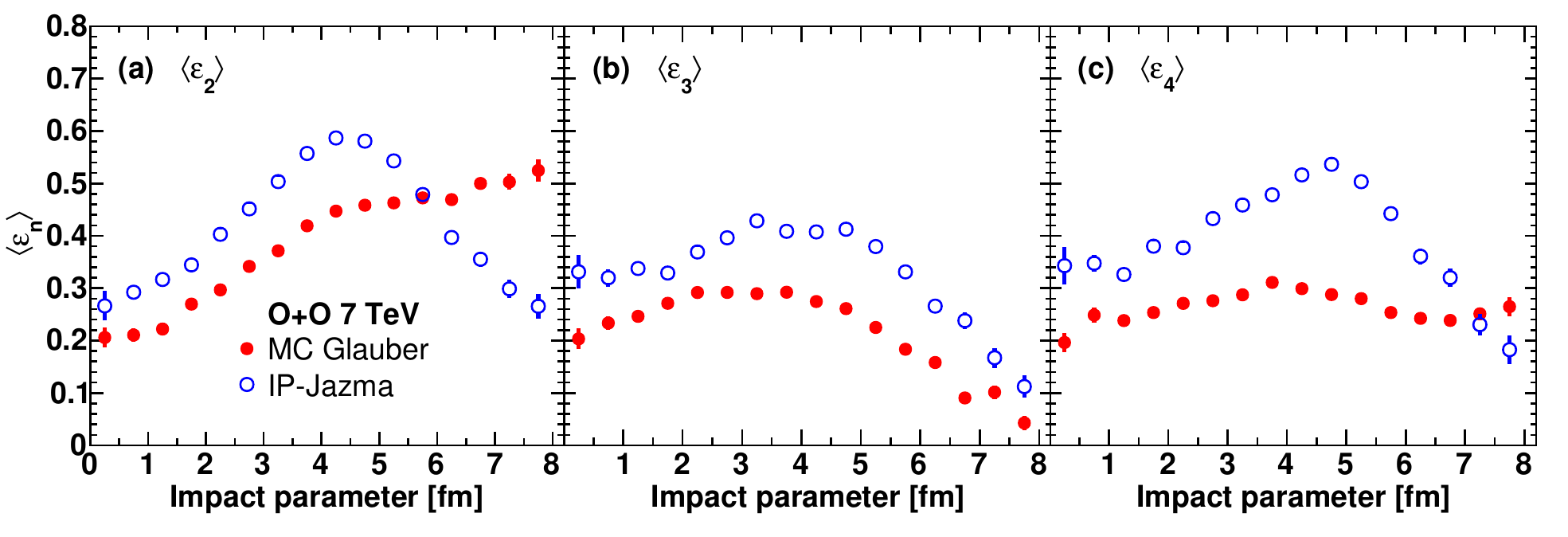}
\caption{\label{fig:ecc_OO_mcglauber_ipjazma}
Comparison of the mean eccentricity as a function of impact parameter in \OO collisions between MC Glauber and IP-Jazma initial conditions.}
\end{figure*}

\begin{figure*}[htb]
\includegraphics[width=1.00\linewidth]{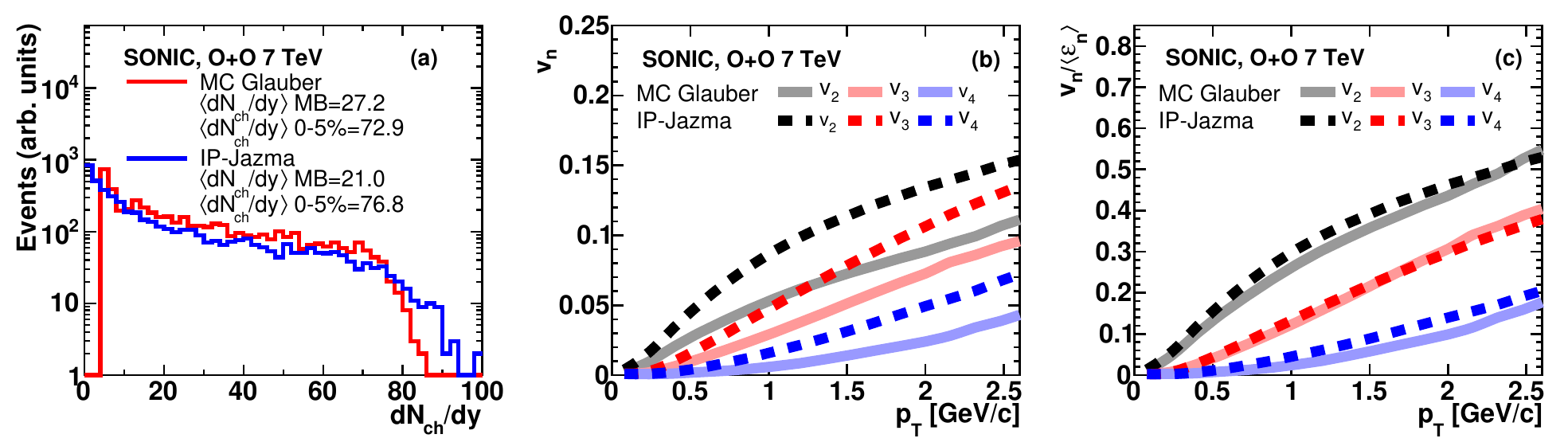}
\caption{\label{fig:vn_pt_OO_mcglauber_ipjazma}
Comparison of $dN_{ch}/d\eta$, $v_n$, and $v_{n}/\langle\varepsilon_{n}\rangle$ in \OO collisions at \sqsn=~7 TeV from \sonic between MC Glauber and IP-Jazma initial conditions.}
\end{figure*}

We begin with results of calculations for \pO and \OO collisions at \sqsn=~7~TeV at the LHC.    Shown in Fig.~\ref{fig:ecc_pO_OO} 
are the mean initial geometric eccentricities ($\langle\varepsilon_{2}\rangle,~\langle\varepsilon_{3}\rangle,~\langle\varepsilon_{4}\rangle$) as a function
of the collision impact parameter.    These results are obtained using the Monte Carlo Glauber initial conditions with a
simple nucleon participant Gaussian smearing.   The results have somewhat similar 
$\langle\varepsilon_{2}\rangle \approx \langle\varepsilon_{3}\rangle \approx \langle\varepsilon_{4}\rangle$ in the small impact parameter range ($<2$ fm) as expected if there is very little intrinsic
geometric shape and the moments are dominated by fluctuations.




A full ensemble of minimum bias events are run through the \sonic calculation with an energy density scale factor chosen to match the LHC \ppb data. The resulting \dnchdeta at midrapidity in \pO (left) and \OO (right) are shown in Figure~\ref{fig:mult_sonic}.
An example \OO event is shown in Figure~\ref{fig:event_OO} stepping from the initial geometry through the time
evolution.    Because the geometry is dominated by fluctuations, each event looks quite distinct, and because the
number of nucleons is small there are quite wide variations.   In this particular event, the initial condition has a somewhat
distinct quadrangular shape (i.e., large $\varepsilon_{4}$) and after evolving hydrodynamically to $t=4$~fm/c there
are four outward patterns corresponding to a larger push along the short axes of the initial geometry (i.e., along the directions of steepest density gradient).

When averaging over many such events, we plot the flow coefficients as a function of \pt in Fig.~\ref{fig:vn_pt_pO_OO_pPb_0005} for events in the highest 5\% multiplicity class of
\pO, \OO at \sqsn=~7 TeV and \pPb at \sqsn=~8.16~TeV.
In the upper panels are the flow coefficients $v_{2}$, $v_{3}$, and $v_{4}$ from left to right, and in the 
lower panels, taking the ratio relative to the initial geometry eccentricities, are $v_{n}/\langle\varepsilon_{n}\rangle$.
The lower panels thus quantify the translation of the geometry into momentum anisotropies and we find that
for the highest 5\% multiplicity collisions, the largest translation occurs for \OO, then \ppb, and is smallest for \pO.    Of course, this follows the relative multiplicities \mdnchdeta going from 72.9, 41.8, and
18.2 respectively.  In Fig.~\ref{fig:vn_pt_pO_OO_pPb_match}, if we specifically select event categories such that all three systems have the
identical multiplicity of $\mdnchdeta \approx 18$ the ordering changes such that $v_{n}/\langle\varepsilon_{n}\rangle$
is approximately the same for \pO and \pPb, and smaller for \OO.     The \OO geometry in this multiplicity
class has the energy density spread over a larger area.   The \pO and \pPb geometries are more compact and thus have a steeper pressure gradient that translates into a slightly larger momentum anisotropy.

For exploring the initial condition dependence, we have run the identical \sonic hydrodynamic evolution code on IP-Jazma generated initial
conditions for \OO collisions.    Figure~\ref{fig:ecc_OO_mcglauber_ipjazma} shows the initial geometry eccentricities comparing the Monte Carlo Glauber and IP-Jazma results.    In the small impact parameter collisions, the IP-Jazma initial conditions results in significantly larger eccentricities which is
expected because the ``hot spots'' will be smaller because the energy deposit is a multiplicative result from the projectile and target color charge
distributions.    An interesting feature is that at large impact parameter, the IP-Jazma eccentricities all tend towards zero.    In the case
of a single nucleon-nucleon collision, the multiplication of two Gaussian color charge distributions, i.e., one from each nucleon, yields exactly a Gaussian which is circularly symmetric and has $\varepsilon_{n} = 0$.   We note that these eccentricities in IP-Jazma are sensitive to the
IP-Sat Gaussian width and a value larger than 0.32~fm as used here will reduce the eccentricities.   A value of 0.50~fm
reduces the $\varepsilon_{2}$ to the same level as the Monte Carlo Glauber for 
$b < 5$~fm.

Figure~\ref{fig:vn_pt_OO_mcglauber_ipjazma} shows a comparison between the previously discussed Monte Carlo Glauber initial conditions and the new IP-Jazma initial conditions both run through the \sonic evolution.    The left panel compares the multiplicity distributions in both cases.    Note that at the lowest multiplicity, the Monte Carlo Glauber case cuts off because one requires at least one nucleon-nucleon collision whereas the 
IP-Jazma calculation can have a smaller and smaller overlap even in the one nucleon-nucleon collision case.    The middle and right panels
compare the resulting momentum anisotropies $v_{n}$ and the anisotropies relative to the initial eccentricity $v_{n}/\langle\varepsilon_{n}\rangle$, 
respectively.    As expected, the $v_{n}$ are larger with the IP-Jazma initial conditions but most of this effect is accounted for when
normalizing by the initial geometric deformation.    As seen when comparing different systems, the IP-Jazma initial hot spots are smaller
and thus have larger pressure gradients and thus after evolution have a slightly larger $v_{n}/\langle\varepsilon_{n}\rangle$.    Full discrimination
between different initial conditions will require the simultaneous constraints from experimental data on the multiplicity distributions,
particle \pt spectra, and flow coefficients.

\newpage

\section{Predictions:  Four ($^4$He) is better than three ($^{3}$He)}

\begin{figure*}[htb]
\includegraphics[width=1.00\linewidth]{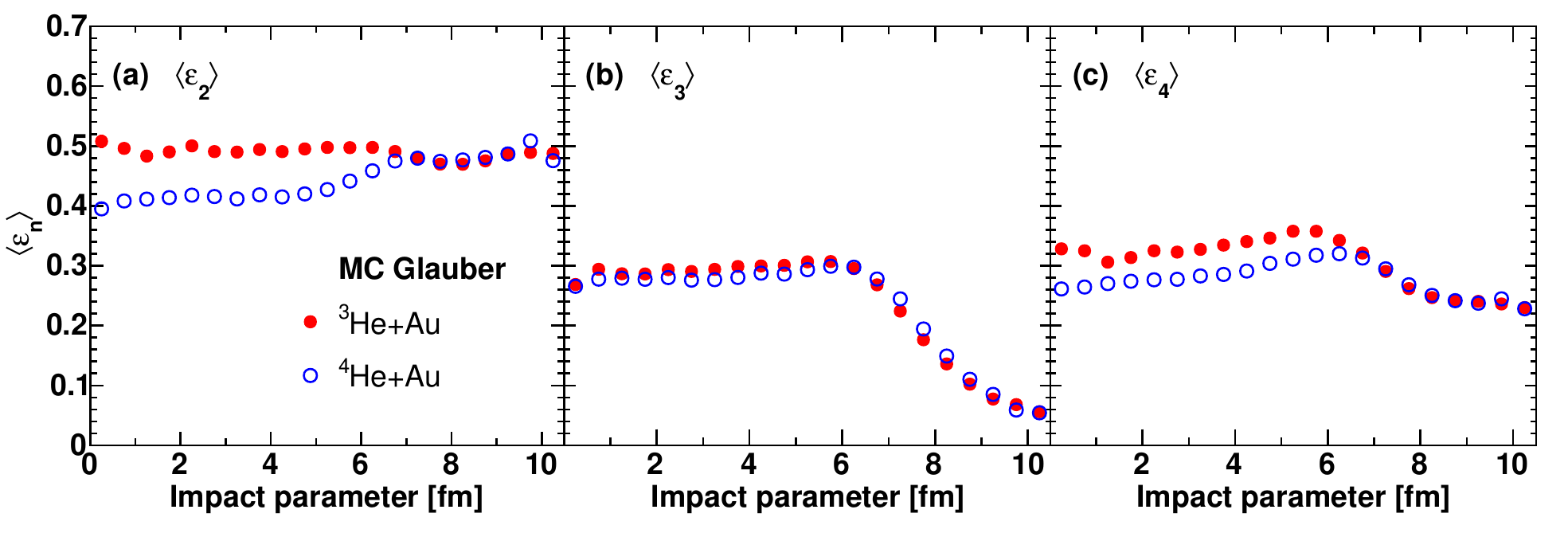}
\caption{\label{fig:ecc_HeAu}
Mean eccentricity ($\langle\ecc_{2}\rangle,~\langle\ecc_{3}\rangle,~\langle\ecc_{4}\rangle$) as a function of impact parameter in \heauthree and \heaufour collisions at $\sqsn=200~{\rm GeV}$.}
\end{figure*}

\begin{figure*}[htb]
\includegraphics[width=1.00\linewidth]{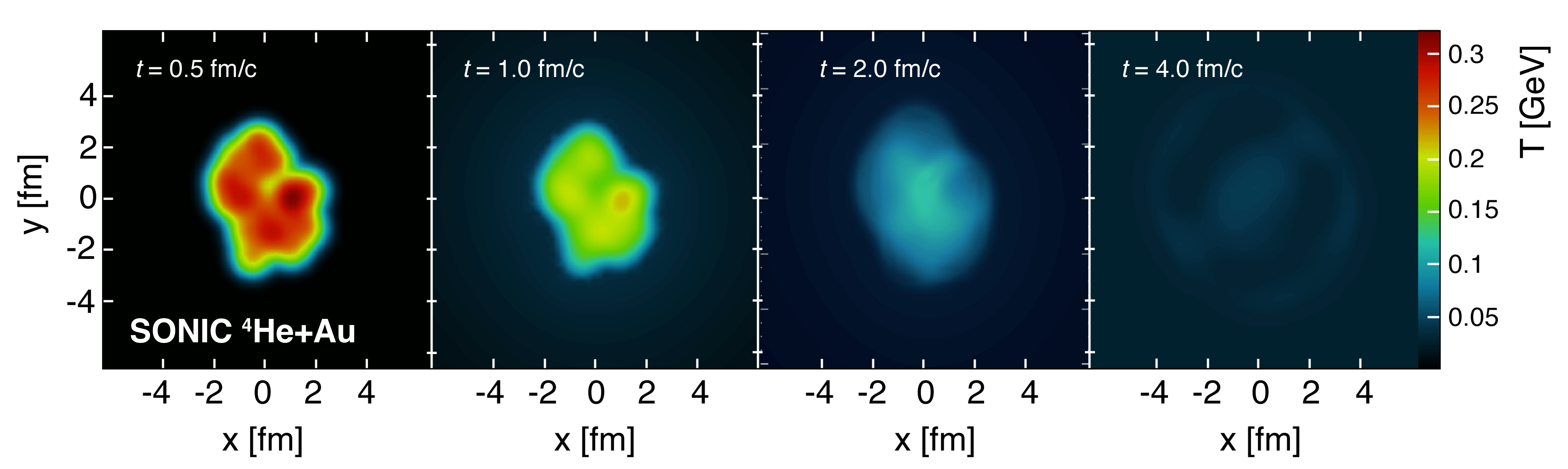}
\caption{\label{fig:event_HeAu}
An example of time evolution of a \heaufour event from \sonic; the color scale indicates the local temperature.}
\end{figure*}

\begin{figure*}[htb]
\includegraphics[width=1.00\linewidth]{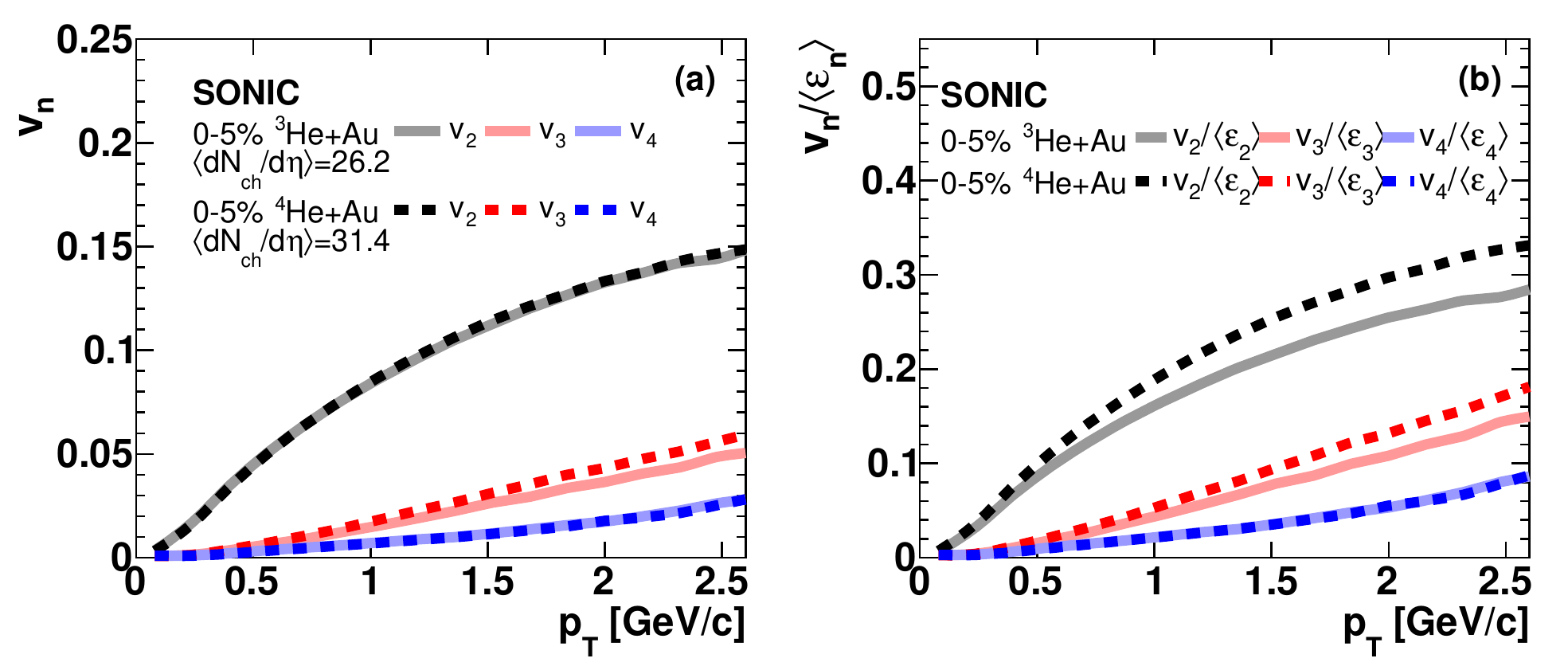}
\caption{\label{fig:vn_pt_HeAu_0005}
Comparison of $v_n$ and $v_{n}/\langle\ecc_{n}\rangle$ from \sonic as a function of \pt in 0\%--5\% of \heauthree and \heaufour collisions at $\sqsn=200~{\rm GeV}$.}
\end{figure*}

\begin{figure*}[htb]
\includegraphics[width=1.00\linewidth]{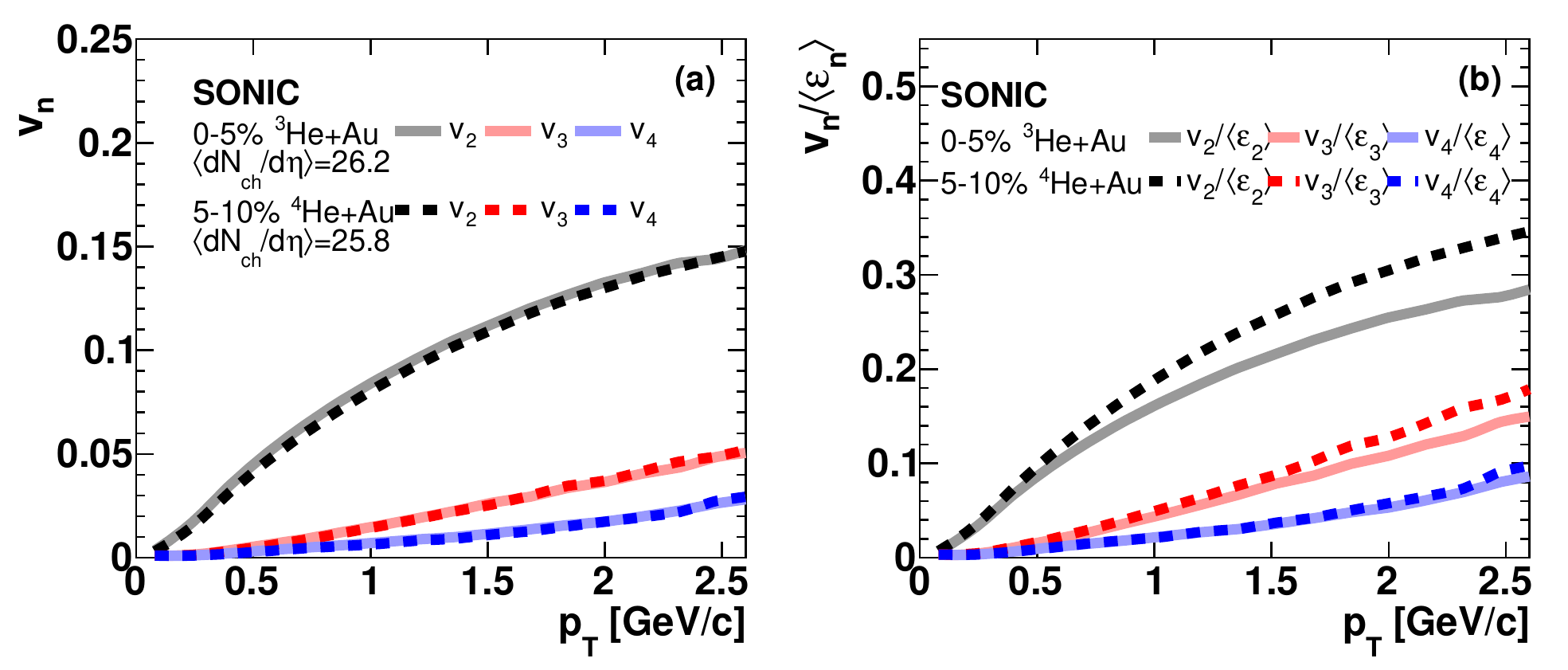}
\caption{\label{fig:vn_pt_HeAu_match}
Comparison of $v_n$ and $v_{n}/\langle\ecc_{n}\rangle$ from \sonic as a function of \pt in 0\%--5\% of \heauthree and 5\%--10\% of \heaufour collisions at $\sqsn=200~{\rm GeV}$.}
\end{figure*}

For the comparison of \heauthree and \heaufour, we first compute the root-mean-square (RMS) value for the 
three-dimensional nucleon central coordinates for $^{3}$He and $^{4}$He which are 1.57~fm and 1.46~fm, respectively.    Hence the $^{4}$He, being more tightly bound, does in fact have its nucleons most closely held together spatially.
Note that these values should not be compared with the RMS charge radius or RMS matter radius for these nuclei,
as the quoted values are just accounting for the central coordinate of each nucleon.    The full calculations detailed above have been checked and are in good agreement with the measured charge radius.   Monte Carlo Glauber results for \heauthree and \heaufour at \sqsn=~200 GeV using the full three- and four- nucleon configurations are shown
in Fig.~\ref{fig:ecc_HeAu} characterizing the eccentricities as a function of impact parameter.   The eccentricities are similar with larger values for \heauthree, and notably the triangularity $\langle\varepsilon_{3}\rangle$ is quite similar
in the two collision systems.

We then run a full ensemble of minimum bias collisions for both systems using identical geometry modeling and
hydrodynamic \sonic parameters.    An example event for \heaufour is shown in Fig.~\ref{fig:event_HeAu}.  
We then calculate the flow coefficients as a function of \pt for both systems in the highest 5\% multiplicity
events as shown in Fig.~\ref{fig:vn_pt_HeAu_0005}.   The resulting $v_{n}$ values are quite similar for the
two systems, though the translation from geometry as characterized by $v_{n}/\langle\varepsilon_{n}\rangle$ is larger for the \heaufour system.    To determine if this is related to the slightly higher multiplicity or the more compact
initial geometry, we make the comparison in Fig.~\ref{fig:vn_pt_HeAu_match} where the event categories
are selected to match in multiplicity.    The results confirm that, just as in the \pO, \OO, and \pPb comparison above,
it is also true in the \heauthree and \heaufour case that the more compact source leads to larger flow.

\newpage

\section{Carbon/Oxygen/Beryllium and Alpha-Clusterization}

Numerous studies have been carried out following the hypothesis that some light nuclei have structures dominated by
$\alpha$ clustering~\cite{Rybczynski:2017nrx,Zhang:2017xda,Bozek:2014cva}.  For example, if a carbon nucleus was composed of three $\alpha$ clusters centered on the corners of a triangle, observable consequences might be seen in the measurement of enhanced triangular flow in collisions of carbon on a heavy nucleus, e.g., \cau.   Of course, if the goal is to resolve questions regarding nuclear structure, the effects must be significantly larger than uncertainties in the hydrodynamic evolution translating these initial structures into momentum flow anisotropies.   Most analyses have implemented very simple nuclear geometries, and we have implemented one such geometry following Ref.~\cite{Rybczynski:2017nrx}.   
For carbon, one assumes that three $\alpha$-like clusters are centered at the corners of a perfect equilateral
triangle with the length of each side taken to be $L=2.8$~fm.    The RMS distance of each nucleon within
the $\alpha$-like cluster relative to a corner of the triangle is set to $r_{c}=1.1$~fm.   
This is significantly smaller than the RMS radius of 1.46~fm for nucleon center coordinates in the $^{4}$He configurations discussed above.   In a similar fashion, Oxygen is assumed to have $\alpha$-like clusters at the corners of a tetrahedron---a pyramid with four sides, each being an identical equilateral triangle.   The length along each triangle is set to $L=3.2$~fm, and again the RMS distance of each nucleon within the $\alpha$-like cluster relative to a corner of the triangle is set to $r_{c}=1.1$~fm.    It is notable that in some other works, for example, Ref.~\cite{Bozek:2014cva}, they also enforce a minimum closest distance between nucleons of $d=0.9$~fm.   In a recent
analysis of \ppb and \pbpb data, a similarly large value of $d=1.13$~fm was favored though with large
uncertainties~\cite{Moreland:2018gsh}.    We do not include any minimum distance and highlight that 
in the full $N$-nucleon configurations there nucleon pairs at significantly smaller separations and thus
these large $d$ values are likely compensating for some other geometric effect.

\begin{figure*}[htb]
\includegraphics[width=\linewidth]{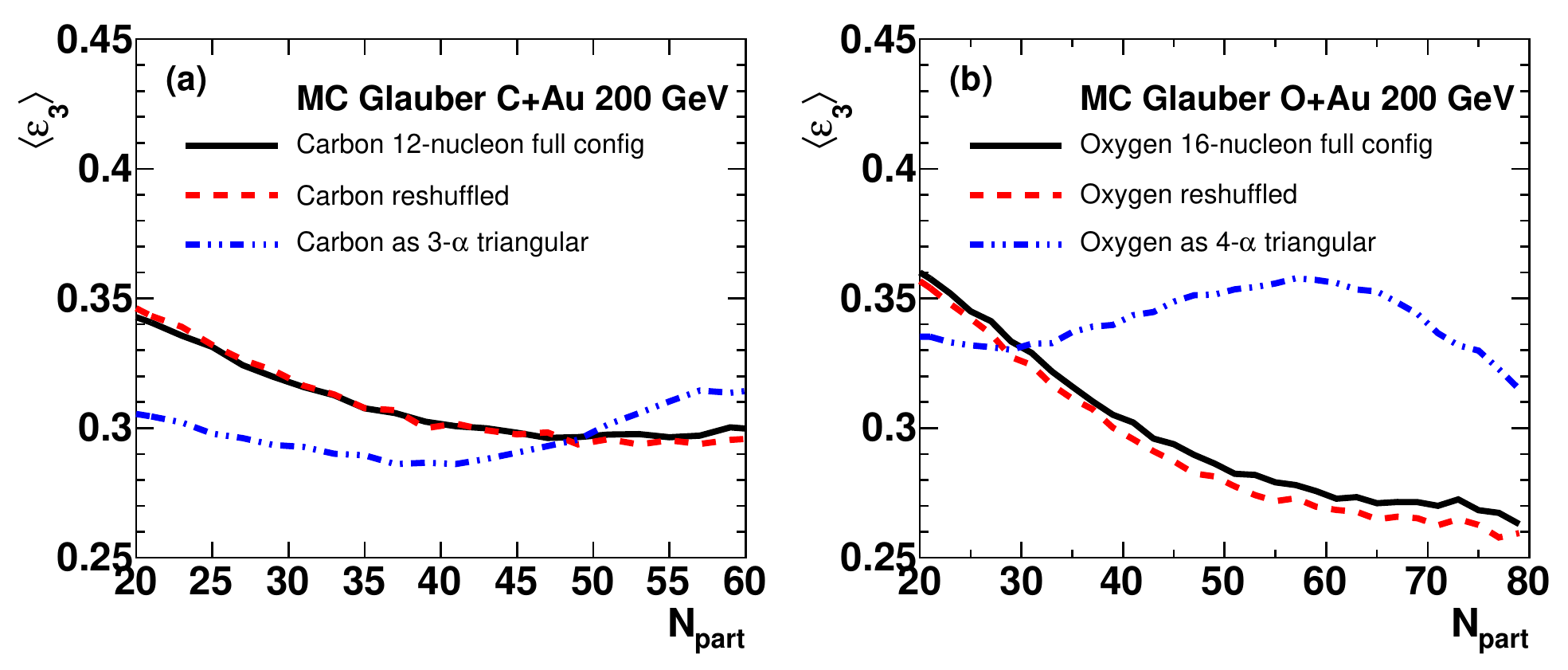}
\caption{\label{fig:triangles}
Spatial triangularity $\langle\varepsilon_{3}\rangle$ is shown as a function of number of nucleon participants for \cau (left) and
\oau (right) collisions at \sqsn=200 GeV.    Results are shown utilizing the full 12- and 16- nucleon configurations 
(black), the reshuffled nucleon configurations with no correlations (red), and with the toy geometry model involving simple triangles and tetrahedra (blue).
}
\end{figure*}

Figure~\ref{fig:triangles} shows the $\langle\varepsilon_{3}\rangle$ as a function of the number of participating nucleons from Monte Carlo Glauber simulations of \cau (left) and \oau (right).   The blue lines are the results from the above detailed triangle and tetrahedron configurations for carbon and oxygen, respectively.   In the \cau case, for $\npart>45$ there is an increase in the triangularity, and then for $\npart>65$ a slight decrease is seen.    The intrinsic triangularity imposed by the carbon geometry is substantially washed out by the fact that the plane of the triangle is randomly oriented in three-dimensional space.   However, the number of participating nucleons is increased by having the triangle oriented in the transverse plane such that the three $\alpha$ clusters hit different parts of the target nucleus.    This explains the enhancement.    Once the triangle is oriented as such, one can gain some further increase in hitting the target nucleons by having the randomly thrown nucleons from each $\alpha$ cluster being further spread out from each other, which slightly decreases the triangularity.    If instead of using the parameter
$r_{c}=1.1$~fm, we actually generate the four-nucleon positions relative to the corner coordinate by sampling from the full $^{4}$He configurations, it is almost equivalent to just using a larger value for $r_{c}$.   We find that
doing so results in a significant reduction in the $\langle\varepsilon_{3}\rangle$ for $\npart>45$.   


A similar pattern for $\langle\varepsilon_{3}\rangle$ as a function of participating nucleons is seen for \oau collisions, 
except that the enhancement is more pronounced even at lower \npart.   This is because the tetrahedron geometry always has some triangularity in the transverse plane regardless of its orientation in three dimensions.

Also shown as black curves are the results from the full 12- and 16- nucleon configurations from carbon and oxygen respectively.    In this case, no strong enhancement is observed.   To glean if any effect of $\alpha$-clustering or nucleon correlations is apparent, we generated so-called ``reshuffled'' configurations such that the radial distribution of nucleons is identical to the full configurations but with no nucleon-nucleon correlations possible.   These reshuffled results are shown as red curves.    One can see only a very small enhancement in the triangularity in both the \cau and \oau cases in the full configuration relative to the reshuffled one.

These results indicate that though there may be some $\alpha$ clustering in full configurations for carbon and oxygen, it is less than indicated in the simple toy geometry picture.   This is not surprising as the toy model result is also seen to be reduced by additional spreading of the cluster geometry $r_{c}$ and it is obvious that there would be event-by-event variations in the triangle configuration parameter $L$.   

\begin{figure*}[htb]
\includegraphics[width=\linewidth]{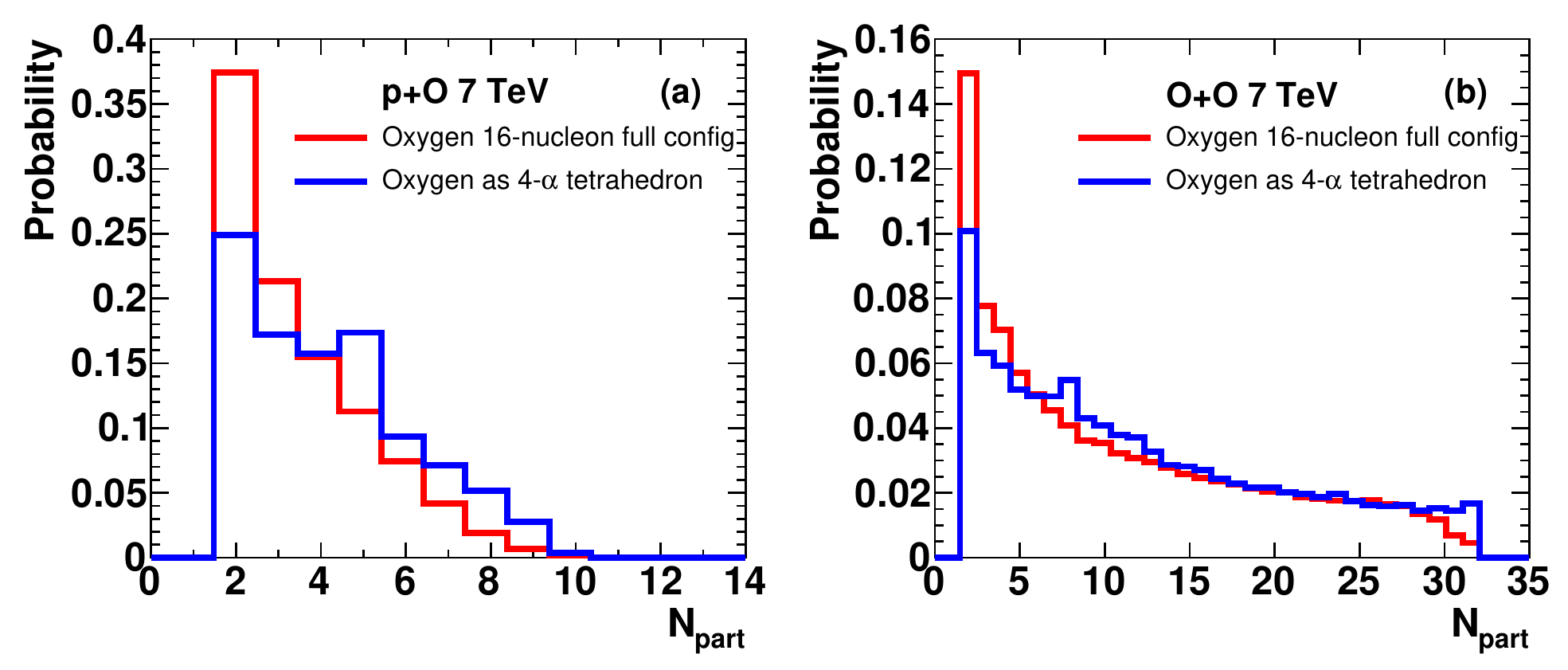}
\caption{\label{fig:alphageomnpart}
Number of participating nucleons in minimum bias \pO (left) and \OO (right) collisions from Monte Carlo Glauber calculations.
Shown are the results of two oxygen cases, one with the full 16-nucleon configurations and one with the
simple $\alpha$-cluster tetrahedron configuration.
}
\end{figure*}

Because the LHC is planning \pO and \OO runs~\cite{Citron:2018lsq}, it is insightful to compare just the distribution of \npart when using the full 16-nucleon configuration for the oxygen as opposed to the $\alpha$-cluster tetrahedron model.   The 
results for \pO (left) and \OO (right) shown in Fig.~\ref{fig:alphageomnpart} show a significant change in the distributions, and much more so for \pO.    In fact, there is a pronounced enhancement in the tetrahedron model in the probability for $\npart = 5$ which means the single projectile proton is striking all four target nucleons, from the
same $\alpha$ cluster.   There is also a significantly larger tail of large \npart values going up to $\npart=9$ 
which correspond to cases where the projectile proton is lined up with one edge of the tetrahedron and thus has
an increased probability to hit all eight nucleons in two $\alpha$ clusters centered on the corners of that edge.
Even though detailed modeling of particle production will be required, the experimental data should easily be able to discriminate the full configuration case from the simple tetrahedron
case.

\begin{figure*}[htb]
\includegraphics[width=\linewidth]{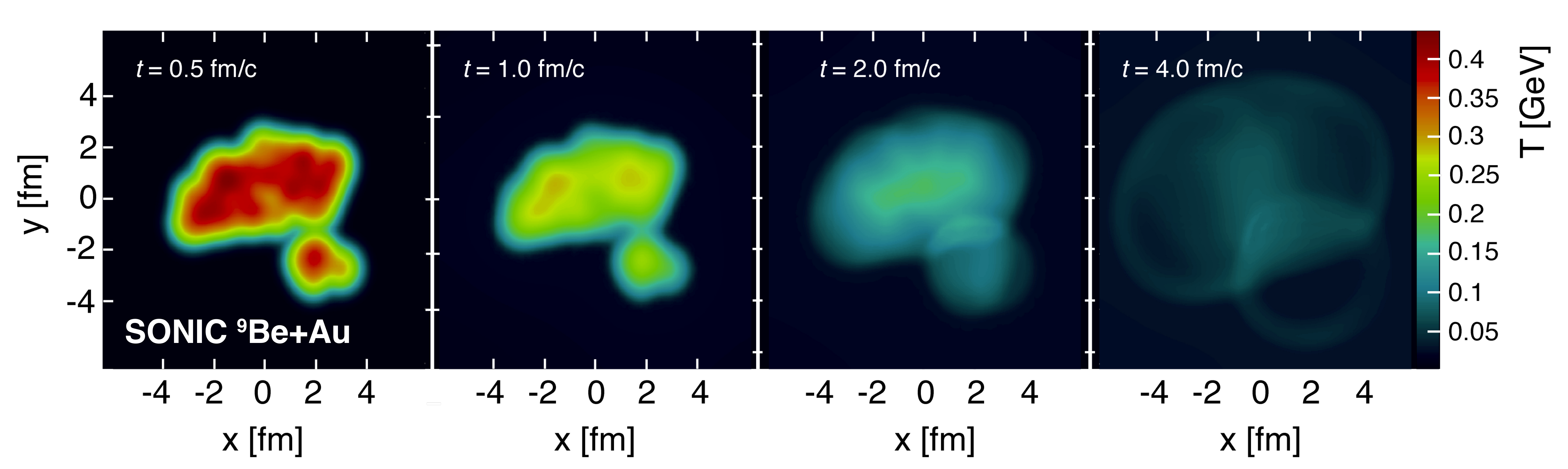}
\caption{\label{fig:be9auevent}
An example of time evolution of a $^{9}$Be$+$Au event from \sonic, and the color scale indicates the local temperature.
In this event, one can see the distinct hot spots from the two $\alpha$ clusters and the extra neutron from 
the $^{9}$Be projectile.
}
\end{figure*}

A potentially more interesting collision projectile, instead of carbon or oxygen on a heavy nucleus, would involve beryllium isotopes $^{7}$Be or $^{9}$Be, as discussed also in Ref.~\cite{Rybczynski:2017nrx}.    The $\alpha$-cluster picture would predict $^{7}$Be as a molecule-like state of $^{4}$He and $^{3}$He, while $^{9}$Be would have two $^{4}$He nuclei and an extra neutron.
This would lead to a significant elliptical deformation $\langle\varepsilon_{2}\rangle$ compared with a spherically symmetric
distribution of seven or nine nucleons.    In this case, there are strong theoretical indications for this deformation~\cite{PhysRevC.83.034312}, though currently there are no available seven- and nine- nucleon configurations, as we have used for other light nuclei.
Such calculations would provide a key input to motivate such collisions of beryllium isotopes on heavy nuclear targets.   One could view this as analogous to running \dau but with a larger energy deposition and over a larger
area for each of the two hot spots.   
An example $^{9}$Be$+$Au event run through the \sonic evolution is shown in Fig.~\ref{fig:be9auevent}.
Running with $^{9}$Be projectiles at RHIC should be possible, while $^{7}$Be would present much greater challenges\footnote{Private communication from Wolfram Fischer}.   
In our view, this would be quite interesting if the nuclear structure were well constrained, thus providing an additional key test of the time evolution dynamics.   
Running a new geometry to test both the nuclear
structure and the evolution would be suboptimal.
Additionally, any such running would require detector configurations capable of measuring long-range correlations via multiple pseudorapidity ranges to disentangle flow and non-flow.

\section{Summary}

Running different nuclear collision geometries at RHIC and the LHC has proven 
insightful for understanding the basic physics at play, particularly in small systems.
In this paper we explore additional geometries and make specific predictions using
the publicly available Monte Carlo Glauber, IP-Jazma, and hydrodynamic evolution \sonic codes.
The results incorporate new calculations of the full $N$-nucleon configurations
for $^{4}$He, carbon, and oxygen nuclei.
Hydrodynamic calculations with \sonic indicate flow anisotropies that 
approximately scale with initial eccentricities, and have an additional dependence on the compactness of the initial geometry.
Having robust calculations on the nucleon configurations for
beryllium isotopes could make \beau collisions of great interest.  
Additionally, a future upgrade at the LHC to allow
collision between two different nuclei would be most beneficial.   

\section*{Acknowledgements}
We acknowledge Paul Romatschke for useful discussions, support of the publicly available \sonic code, and a careful reading of the manuscript.   We acknowledge useful discussions on running options at RHIC with Wolfram Fischer in Brookhaven National Laboratory, Collider-Accelerator Division.   We acknowledge Stefano Gandolfi for his help and support in generating the $^{16}$O configurations from quantum Monte Carlo simulations.   J.L.N. also acknowledges Giuliano Giacalone, Jean-Yves Ollitrault, and Hugo Pereira da Costa for many helpful discussions on small system geometry. 

S.H.L., J.L.N., and J.O.K. acknowledge support from the U.S. Department of Energy, Office of Science, Office of Nuclear Physics under Contract No. DE-FG02-00ER41152.  J.L.N. is thankful for generous support from
CEA/IPhT/Saclay during his sabbatical time in France and the Munich Institute for Astro- and Particle Physics (MIAPP) 
of the DFG cluster of excellence ``Origin and Structure of the Universe'' during his time in Germany.
J.O. acknowledges support from the U.S. Department of Energy, Office of Science, Office of Nuclear
Physics under Contract No. DE-SC0018117.
C.L. is supported by the U.S. Department of Energy, Office of Science, Office of Nuclear Physics, under Contract No. DE-AC05-00OR22725.
The work of D.L. was supported by the U.S. Department of Energy, Office of Science, Office of Nuclear Physics, under Contract No. DE-SC0013617, and by the NUCLEI SciDAC program. 
Computational resources have been provided by Los Alamos Open Supercomputing via the Institutional Computing (IC) program, and by the National Energy Research Scientific Computing Center (NERSC), which is supported by the U.S. Department of Energy, Office of Science, under Contract No. DE-AC02-05CH11231. The work of J.E.L. was supported by the BMBF under contract No. 05P15RDFN1. Calculations for this research were also conducted on the Lichtenberg high performance computer of the TU Darmstadt.

\clearpage
\appendix

\section{Predictions from the \textsc{AMPT} Model}

In the hydrodynamic paradigm, small system collectivity arises from the translation of the initial collision geometry into the final state through interactions among fluid elements. In contrast, the model known as A-Multi-Phase-Transport~\cite{Lin:2004en} (\textsc{ampt})
achieves this same translation in its string melting mode by modeling the medium as a collection of interacting quarks and antiquarks. In brief, the model uses \textsc{hijing}~\cite{Wang:1991hta} for the geometric initialization of the colliding nuclei, extending Lund color strings between pairs of interacting nucleons from which hadrons are emitted. These hadrons then dissociate into their valence quark content, which enter a partonic scattering stage. At the end of this stage, hadronization proceeds via coalescence of (anti)quarks in close spatial proximity. Lastly, the hadrons enter a hadronic scattering stage.

\begin{figure}[hbt]
    \includegraphics[width=\linewidth]{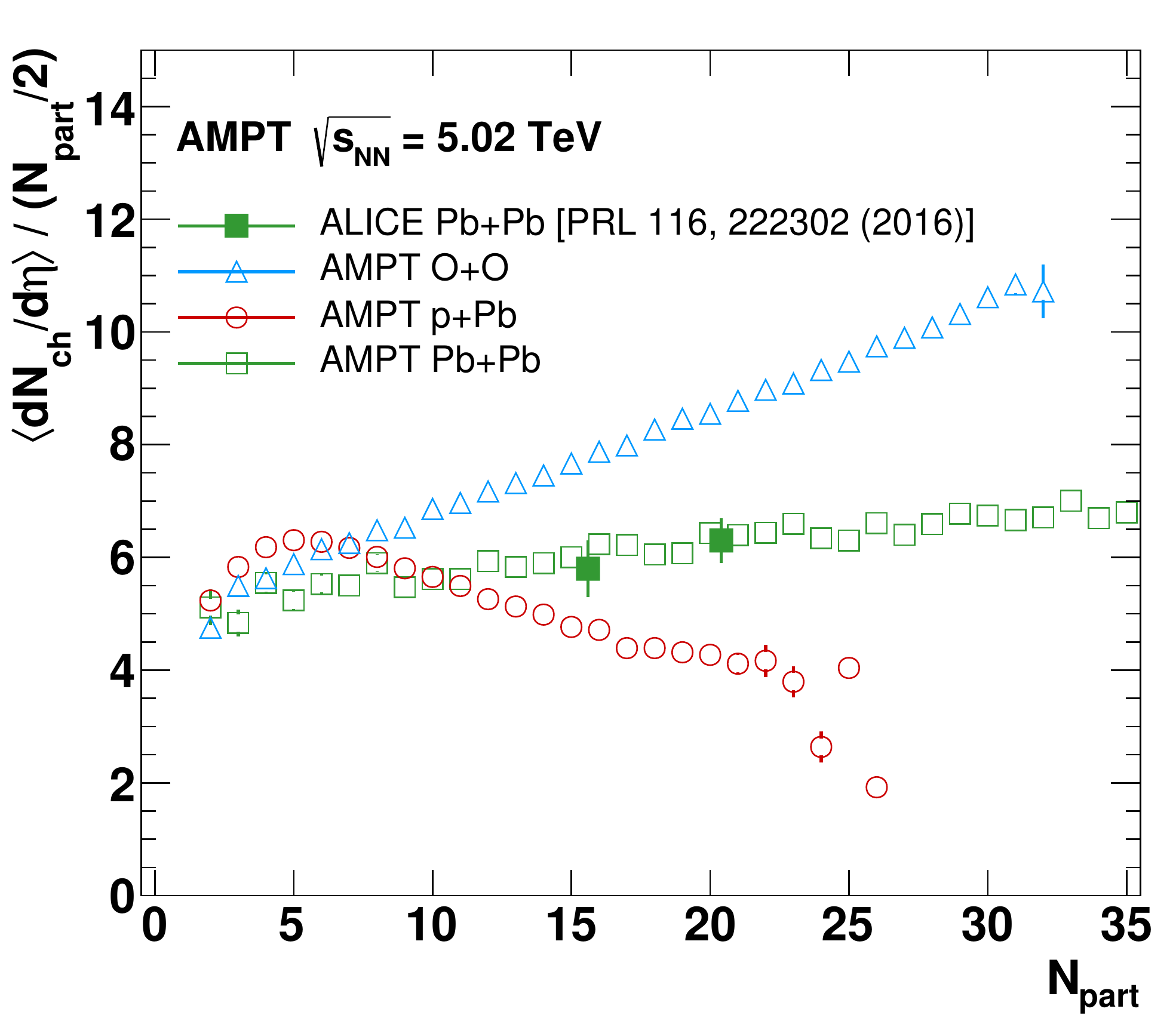}
    \caption{Mean charged particle production per pair participant ($\mdnchdeta / ({N_{\rm part}}/2)$) for \pPb, \pbpb, and \OO collisions in \textsc{ampt} at $\sqrt{s_{NN}}=5.02$~TeV with the same Lund string fragmentation function parameters ($a = 0.3$, $b=0.15$). Measurements by the ALICE collaboration in \pbpb collisions~\cite{Adam:2015ptt} are shown for comparison.}
    \label{fig_ampt_particleprod}
\end{figure}

\begin{figure*}[htb]
\includegraphics[width=1.00\linewidth]{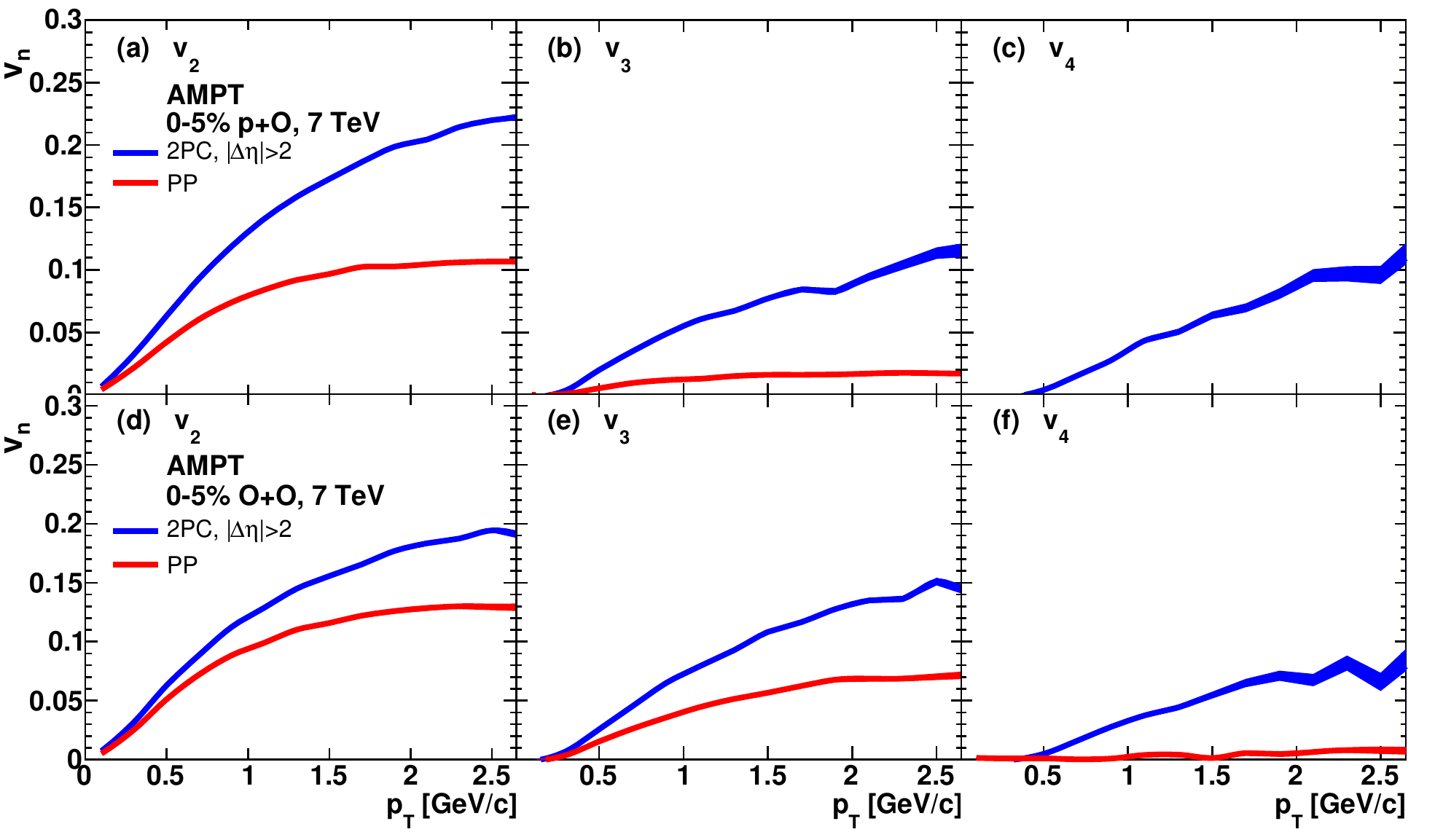}
\caption{\label{fig:vn_pt_pO_OO_ampt}
Flow coefficients $v_n$ as a function of \pt in the highest 0--5\% centrality class in \pO (a,b,c) and \OO (d,e,f) collisions at \sqsn=~7~TeV, as calculated with the \ampt model. Two calculations are presented, namely the true flow $v_n\{\text{PP}\}$ relative to the parton participant plane, and the two-particle correlation flow $v_n\{\text{2PC}\}$.}
\end{figure*}

The \ampt model was successfully used to describe various aspects of small system collectivity, as observed both at LHC~\cite{Bzdak:2014dia} and RHIC energies~\cite{Koop:2015wea, Koop:2015trj, Aidala:2017pup, Aidala:2017ajz}. It is, therefore, of interest to examine flow predictions from the publicly available \ampt model in \pO and \OO collisions.

Final-state particle production in \ampt depends primarily on the parametrization of the Lund string fragmentation function, which determines the tension of the color string. It was shown that a specific set of string parameters (namely, $a=0.3$ and $b=0.15$ GeV$^{-2}$) are well suited to simultaneously describe charged particle multiplicity, spectra, and particle-identified $v_2$ in \pbpb and \pPb collisions at LHC energies~\cite{Lin:2014tya}. 
In the absence of experimental \pO and \OO data to constrain the model, we take the above set of string parameters to run our \ampt calculations. All \ampt results are calculated with a parton-parton cross section
$\sigma=3$~mb.

Figure~\ref{fig_ampt_particleprod} shows particle production at midrapidity per participant nucleon pair,  $\mdnchdeta / ({N_{\rm part}}/2)$, as obtained from \ampt for \pbpb, \ppb, and \OO collisions at $\sqrt{s_{NN}}=5.02$~TeV.   Here we show all systems at the same $\sqrt{s_{NN}}$ to directly compare particle
production.  Shown for comparison is a corresponding experimental measurement in \pbpb by the ALICE collaboration~\cite{Adam:2015ptt}, which demonstrates that the chosen string parameters indeed allow the model to reproduce particle production as observed in \pbpb collisions at LHC energies.

A striking feature of Fig.~\ref{fig_ampt_particleprod} is the steep rise in particle production in \OO compared to \pbpb collisions.   It is notable that the ratio of the number of binary collisions \ncoll to number of 
participating nucleons \npart is significantly higher in \OO collisions compared to either \pbpb or \ppb collisions and it may explain the difference in particle production scaling in \ampt.   We highlight that in the
hydrodynamic \sonic calculations shown earlier, we follow the assumption that deposited energy density is
additive via the contributions from each participating nucleon.   Thus, the \sonic calculations have a significantly
lower \mdnchdeta in the 0--5\% highest multiplicity events compared with the \textsc{ampt} results.    Experimental data will be needed to constrain these particle production models before a final comparison
can be made with the calculated flow coefficients.


Figure~\ref{fig:vn_pt_pO_OO_ampt} shows flow coefficients $v_2$, $v_3$, and $v_4$ for unidentified charged hadrons in \ampt \pO and \OO collisions at $\sqrt{s_{NN}}=7$~TeV, as a function of particle $p_T$ for the 0\%--5\% most central events. Flow coefficients are calculated in two different ways. The \textit{true flow}, defined as that relative to the initial event geometry, is calculated as follows 
\begin{equation}
    v_n\{\text{PP}\} = \langle \cos[n(\phi-\Psi_n)]  \rangle,
\end{equation}
where the collision plane $\Psi_n$ is estimated using the early-time coordinates of partons as they emerge from string melting. In contrast, flow coefficients calculated using two-particle correlations are given by
\begin{equation}
    v_n^2\{\text{2PC}\} = \langle \exp [in(\phi_1 - \phi_2)] \rangle
\end{equation}
and include non-flow contributions from processes such as jet fragmentation, resonance decays, etc.  The $v_n\{\text{2PC}\}$ are always larger than the corresponding $v_n\{\text{PP}\}$.    One reason is that the
former also include these non-flow contributions that are random with respect to the initial geometry.   
Furthermore, $v_n\{\text{2PC}\}$ is a measure of the RMS of the flow coefficient, while the $v_n\{\text{PP}\}$ is
strictly the average.   This difference is particularly notable in the $v_{4}$ results where the quantity
$v_n\{\text{PP}\}$ has a significant cancellation effect from positive and negative contributions, and the average
in the \pO case is in fact negative (and hence does not appear in the panel).


Both the \ampt $v_2\{\text{2PC}\}$ and $v_2\{\text{PP}\}$ are larger than the \sonic results.   This is 
expected just from the larger particle production in \textsc{ampt}.   One could tune the \textsc{ampt} selection
to more closer match the participant scalings used for the \sonic energy density input; however, we leave this as
a future exercise when experimental data can constrain the models.


\clearpage



\bibliography{main}   

\end{document}